\documentclass[aps,nofootinbib]{revtex4}
\usepackage{amsmath}
\usepackage{graphicx}

\begin{document}

\title{Power counting with one-pion exchange}
\author{Michael C. Birse}
\affiliation{Theoretical Physics Group, School of Physics and Astronomy,\\
The University of Manchester, Manchester, M13 9PL, UK\\}

\begin{abstract}
Techniques developed for handing inverse-power-law potentials in atomic physics
are applied to the tensor one-pion exchange potential to determine the 
regions in which it can be treated perturbatively. In $S$-, $P$- and $D$-waves
the critical values of the relative momentum are less than or of the order of 
400~MeV. The RG is then used to determine the power counting for short-range 
interaction in the presence of this potential. In the $P$-and $D$-waves, where there 
are no low-energy bound or virtual states, these interactions have half-integer RG 
eigenvalues and are substantially promoted relative to naive expectations. These 
results are independent of whether the tensor force is attractive or repulsive. In 
the $^3S_1$ channel the leading term is relevant, but it is demoted by half an order 
compared to the counting for the effective-range expansion with only a short-range 
potential. The tensor force can be treated perturbatively in those $F$-waves and above
that do not couple to $P$- or $D$-waves. The corresponding power counting is
the usual one given by naive dimensional analysis.
\end{abstract}
\maketitle

\vskip 10pt

\section{Introduction}

Since Weinberg \cite{wein} first proposed that the ideas of chiral perturbation 
theory (ChPT) could be applied to nuclear forces, there has been a continuing 
debate over which parts of interaction can be treated perturbatively and which 
can, or indeed must, be treated nonperturbatively. This has led to two widely-used
schemes for constructing effective field theories (EFT's) to represent 
these forces.\footnote{For reviews, see Refs.~\cite{border,bvkrev}.}

One is based on Weinberg's original suggestion \cite{wein} and has been widely
applied by van Kolck and collaborators \cite{bvkrev,orvk,bmpvk}. It will be 
referred to here as the WvK scheme. 
In it one first expands potential using perturbative ``Weinberg" power 
counting (like that in ChPT for mesons or single nucleons \cite{scherer}).
Then one constructs the scattering amplitude by iterating
the lowest-order terms: the leading, energy-independent 
contact interaction and one-pion exchange (OPE).

The other scheme, developed by Kaplan, Savage and Wise (KSW) \cite{ksw}, starts 
from a nontrivial fixed point of the renormalisation group, which corresponds to 
a two-body system with an infinite scattering length. In the expansion around 
this point, all pion-exchange forces as well as momentum- or energy-dependent
contact interactions are treated as perturbations.

At very low momenta, pion-range physics is not resolved and nuclear forces can be 
described just in terms of contact interactions. In this regime the two schemes, 
WvK and KSW, are equivalent, since iterating the leading contact interaction leads 
to the same power counting \cite{bvk,vk,ksw}, and they just reproduce the 
effective-range expansion \cite{bethe,newton}. However at higher momenta, they 
treat OPE differently. There, problems with the perturbative KSW scheme have been
identified by Fleming, Mehen and Stewart \cite{fms}. In particular they have 
shown that the expansion is only slowly convergent in the $^1S_0$ channel and, 
worse, it seems not to converge at all in the $^3S_1$ channel. This has led Beane 
{\em et al.}~\cite{bbsvk} to propose a hybrid approach, using KSW in the former 
and WvK in the latter. 

The strength of the OPE potential is given by the square of the pseudovector 
coupling which, to leading chiral order, is given by 
$f_{\pi{\scriptscriptstyle NN}}^2=g_{\scriptscriptstyle A}^2 m_\pi^2/(16 \pi F_\pi^2)$.
Empirical determinations lead to values for $f_{\pi{\scriptscriptstyle NN}}^2$
of about 0.075 \cite{nijnn}. If we factor the nucleon mass out of the 
Hamiltonian, we find that the OPE potential contains the scale
\begin{equation}
\lambda_\pi=\frac{16\pi F_\pi^2}{g_{\scriptscriptstyle A}^2 
M_{\scriptscriptstyle N}}\simeq 290\;\mbox{MeV},
\label{eq:lambdapi}
\end{equation}
which is constructed from the nucleon mass and the pion decay constant. Since 
both of these are high-energy scales in ChPT, one would naturally take this to 
be a high-energy quantity. If $\lambda_\pi$ were much larger than $m_\pi$, OPE
could the be treated as weak for scattering momenta of the order of $m_\pi$ and the 
KSW scheme would apply. This would be the case in a world with much smaller up- 
and down-quark masses, where one would be much closer to the chiral limit
and $m_\pi$ would be smaller than $f_\pi$, not just smaller than 
$4\pi f_\pi$---the typical combination that appears in chiral expansions for
processes involving at most one nucleon \cite{scherer}.
Unfortunately, in the real world $\lambda_\pi$ is only about twice $m_\pi$, 
and so we do not have a good separation of scales. This is what underlies the
difficulties in building a useful perturbative EFT with pion-exchange forces.

The first question is whether to iterate OPE or not. However, Nogga, Timmermans 
and van Kolck \cite{ntvk} have shown numerically that iterated OPE cannot be 
consistently renormalised if the contact interactions are assigned the orders they 
would have in naive dimensional analysis. A similar observation is also made in 
footnote 5 of Ref.~\cite{pvra3}, but without examining in detail the consequences 
for power counting. Hence, if we choose to iterate OPE, we are forced to address 
a second question: what power counting should we use for the resulting contact 
interactions?

In this work, I address the first question by constructing exact solutions to 
the Schr\"odinger equation and examining them for nonanalytic dependence on the
strength of the OPE potential. In the chiral limit, this potential has a
tensor form proportional to $1/r^3$. Solutions can be obtained using techniques
that have been developed in atomic physics \cite{cava,gao1,gao2}. As in the case
of the simple $1/r^3$ potential studied by Gao \cite{gao2}, the solutions in
each partial wave become nonanalytic in the strength above some critical value 
for the dimensionless product of the momentum and coupling strength. This implies
that the potential must be treated nonperturbatively in this region. 

For the $1/r^3$ in uncoupled partial waves, the critical values have been 
determined by Gao \cite{gao2}. Here I apply these results to the tensor potential 
in coupled waves. Since the strength of OPE is given in terms of $ 1/\lambda_\pi$, 
these results can be converted into a critical value for the relative momentum
in each channel, above which OPE must be treated nonperturbatively. In channels
involving waves with $l\le 2$, these critical momenta are $\lesssim 400$~MeV,
implying that OPE needs to be iterated in them, in agreement
with the observations in Ref.~\cite{fms}. In contrast, the critical momenta 
in channels that involve only waves with $l\ge 3$ are well above 1~GeV and
so OPE can treated perturbatively.

To answer the second question, I use the Wilsonian renormalisation 
group (RG) \cite{wrg} to determine the scale dependence of the interactions between 
two nucleons. In this approach, one imposes a floating cut-off $\Lambda$, lying 
between the low-energy scales of interest and high scales of the underlying interest.
Demanding that the scattering amplitude be independent of the cut-off then
leads to an RG equation for the effective short-range potential describing
the physics that is not resolved at the scale $\Lambda$ \cite{bmr,bb1,bb2}. 

This cut-off could be imposed on a plane-wave basis but, in the presence of a 
known long-range potential, it turns out to be more convenient to work in the basis 
of distorted waves (DW's) of that potential \cite{bb1,bb2}.
For a long-range potential that is singular at the origin, the scaling behaviour, 
and hence the power counting, is controlled by the power-law dependence of the DW's 
for small $r$. In the case of the $1/r^3$ tensor potential, I find that the 
short-range interactions have half-integer RG eigenvalues (anomalous dimensions) and 
so their scaling is quite different from that given by naive dimensional analysis.
These eigenvalues are all positive if the scattering is weak, but are 
smaller than they would be in the absence of the long-range potential.
Compared to the usual power counting \cite{wein}, these interactions are
``promoted" to lower orders in the expansion in small scales and hence are
more important for low-energy scattering than one would naively expect.
This agrees with the numerical observations in Ref.~\cite{ntvk}. Similar
conclusions about the need for a modified power counting are drawn in 
Refs.~\cite{pvra1,pvra2,pvra3}, which use a related DW approach with a 
radial cut-off only. The RG analysis here provides the new power counting.
It also shows that in waves where the tensor force can be treated perturbatively, 
the usual power counting still applies.

\section{Solutions in the chiral limit}

The long-range OPE potential has a central piece
\begin{equation}
V_{\pi {\scriptscriptstyle C}}(r)=\frac{1}{3}\,f_{\pi{\scriptscriptstyle NN}}^2
\frac{e^{-m_\pi r}}{r}\, ({\boldsymbol\sigma}_1\cdot{\boldsymbol\sigma}_2)
({\boldsymbol\tau}_1\cdot{\boldsymbol\tau}_2),
\end{equation}
and a tensor piece
\begin{equation}
V_{\pi {\scriptscriptstyle T}}(r)=\frac{1}{3}\,
\frac{f_{\pi{\scriptscriptstyle NN}}^2}{m_\pi^2}
\left(3+3m_\pi r+m_\pi^2 r^2\right)
\frac{e^{-m_\pi r}}{r^3}\, S_{12} ({\boldsymbol\tau}_1
\cdot{\boldsymbol\tau}_2),
\end{equation}
where $S_{12}=3({\boldsymbol\sigma}_1\cdot\hat{\bf r})
({\boldsymbol\sigma}_2\cdot\hat{\bf r})
-{\boldsymbol\sigma}_1\cdot{\boldsymbol\sigma}_2$.

In the spin-singlet channels only the central, Yukawa potential 
$V_{\pi {\scriptscriptstyle C}}$ contributes. This has a $1/r$ singularity at the 
origin. Even when iterated, this is not sufficient to affect the power counting for 
the short-range interactions \cite{bb1}. However, in the case of the spin-triplet 
channels, we have to deal with the tensor piece of OPE. This behaves like $1/r^3$ at 
short distances and so it is not obvious that it can ever be treated perturbatively. 
If we do iterate this interaction, the resulting nonperturbative short-distance 
physics can alter the power counting. 

In fact a $1/r^3$ singularity is sufficiently short-ranged that waves with
low momenta do not resolve the singularity and their scattering can still be 
treated perturbatively. As discussed below, the critical value of the momentum 
above which nonperturbative behaviour sets in is proportional to the scale 
$\lambda_\pi$ in Eq.~(\ref{eq:lambdapi}), and increases rapidly with the 
orbital angular momentum of the wave. This is because in higher partial waves
the centrifugal barrier ``protects" low-energy waves from probing the 
singularity.

At long distances the potential falls off exponentially as a result of the 
finite pion mass. This ensures that there are no nonanalytic terms in the 
scattering amplitude and so, at very low energies, the effective-range 
expansion \cite{bethe,newton} can be applied to it. However it does not alter 
the singular behaviour at sort distances and so the effects of that can still be 
analysed using the simpler form of the potential in the chiral limit
($m_\pi\rightarrow 0$).\footnote{If $\lambda_\pi$ had been much smaller
than $m_\pi$ then this would not be possible; the exponential fall-off would 
mean that waves did not probe the singularity until their momenta were at least
of order $m_\pi$. As discussed below, the critical momenta obtained in the chiral 
limit are greater than $m_\pi$ in all channels except the $^3S_1$--$^3D_1$. Even
in that the case, the critical momentum is of the order of $m_\pi/2$ and so the
conclusion that perturbation theory breaks down for momenta of the order of $m_\pi$
should still be valid, although the precise value of the momentum at which this 
happens will be different.}

In the chiral limit the tensor interaction has the $1/r^3$ form:
\begin{equation}
V_{\pi {\scriptscriptstyle T}}(r)
=\frac{1}{M_{\scriptscriptstyle N}\lambda_\pi}\,
\frac{1}{r^3}\, S_{12} ({\boldsymbol\tau}_1
\cdot{\boldsymbol\tau}_2).
\end{equation}
This and other singular inverse-power-law potentials also arise for systems in 
atomic physics, where techniques for solving the corresponding Schr\"odinger 
equations have developed.\footnote{For a recent review, see Ref.~\cite{sbcefmr}.
A review of older approaches can be found in Ref.~\cite{fls}.} 

In applications in both nuclear and atomic physics these singular potentials 
should not be regarded ``fundamental". Instead they provide the long-range parts 
of effective interactions. At small separations the finite sizes and structures 
of nucleons or atoms cannot be ignored, and other, short-range interactions are 
important. The solutions to the Schr\"odinger equation with one of these potentials 
can be used as a DW basis for analysing the scale-dependence of the associated 
short-range terms. Provided they are viewed in this way, as pieces of effective 
theories, even attractive singular potentials are meaningful (contrary to 
the comment at the end of Ref.~\cite{gao2}.)

\subsection{Uncoupled channels}

The tensor interaction couples spin-triplet partial waves with $l=j\pm 1$, 
such as $^3S_1$ and $^3D_1$, but not those with $l=j$. I consider the latter first. 
For these, the tensor operator is just $S_{12}=2$ \cite{rs} and the potential has a 
simple $1/r^3$ form. The solutions to the corresponding Schr\"odinger equation can 
be constructed as series expansions in Bessel functions using the method
of Refs.~\cite{cava,gao1,gao2}.\footnote{Other closely related methods exist
for solving equations with inverse-power-law potentials. These are based on 
Laurent \cite{mf,stroff,esp} or Bessel-product expansions \cite{holz,khr}.}
Solutions for the pure $1/r^3$ potential have previously
been obtained by Gao \cite{gao2}, but I recap some of the main features of
the method here before applying it to the coupled waves.

The radial Schr\"odinger equation describing the relative motion of the two
particles has the form
\begin{equation}
-\,\frac{1}{M_{\scriptscriptstyle N}}\left[\frac{{\rm d}^2}{{\rm d}r^2}
+\frac{2}{r}\,\frac{{\rm d}}{{\rm d}r}-\frac{l(l+1)}{r^2}\right]\psi(r)
+\frac{B_3}{r^3}\,\psi(r)=E\psi(r).
\end{equation}
The strength of the potential is $B_3=-6/(M_{\scriptscriptstyle N}\lambda_\pi)$ 
in the uncoupled isospin-singlet waves, such as $^3D_2$ and $^3G_4$, 
and $B_3=2/(M_{\scriptscriptstyle N}\lambda_\pi)$ in the isospin-triplet ones, 
such as $^3P_1$ and $^3F_3$. It is convenient to re-express this equation in 
dimensionless form by introducing the coordinate $x=pr$ and the coupling 
$\kappa=p M_{\scriptscriptstyle N} B_3$, where $p=\sqrt{M_{\scriptscriptstyle N}E}$ 
is the on-shell relative momentum. It is also convenient to put the radial equation 
into a form similar to Bessel's equation by defining $\phi(x)=x^{1/2}\psi(x)$. After 
some rearrangement, the resulting equation is
\begin{equation}
\left[x^2\,\frac{{\rm d}^2}{{\rm d}x^2}
+x\,\frac{{\rm d}}{{\rm d}x}+x^2-\left(l+{\textstyle\frac{1}{2}}
\right)^2\right]\phi(x)=\frac{\kappa}{x}\,\phi(x).
\label{eq:phiode}
\end{equation}
In this form we see that the behaviour of the solutions is controlled by the single
combination of the energy and the coupling strength, $\kappa$. 

This equation can be solved analytically with the aid of the methods in 
Refs.~\cite{cava,gao1,gao2}, by expanding $\phi(x)$ in terms of Bessel functions as
\begin{equation}
\phi(x)=\sum_{n=-\infty}^\infty a_n J_{n+\nu}(x).
\label{eq:phiexp}
\end{equation}
The shift in the order by $\nu$ is needed because the interaction on the 
right-hand side generates secular perturbations that must be resummed
\cite{cava} (see also: Ref.~\cite{boamm}, Sec.~11.1).
Substituting the expansion (\ref{eq:phiexp}) into the equation 
(\ref{eq:phiode}) leads to an infinite set of linear equations for 
the coefficients $a_n$:
\begin{equation}
\left[(n+\nu)^2-\left(l+{\textstyle\frac{1}{2}}\right)^2\right]a_n
=\frac{\kappa}{2(n+\nu+1)}\,a_{n+1}+\frac{\kappa}{2(n+\nu-1)}\,a_{n-1},
\qquad -\infty<n<\infty.
\end{equation}
By introducing $b_n=a_n/(n+\nu)$ and
\begin{equation}
f_l(n+\nu)=2(n+\nu)\left[(n+\nu)^2-\left(l
+{\textstyle\frac{1}{2}}\right)^2\right],
\end{equation}
the linear equations can be put into the more symmetric form
\begin{equation}
\kappa b_{n-1}-f_l(n+\nu)b_n+\kappa b_{n+1}=0\qquad -\infty<n<\infty.
\label{eq:bdiffe}
\end{equation}

Following Refs.~\cite{gao1,gao2} (see also Ref.~\cite{mf}) one can solve 
these equations to get a representation of the ratios of coefficients
in terms of a continued fraction. For positive $n$, I define the ratios
\begin{equation}
R_n=\frac{b_n}{\kappa b_{n-1}}.
\label{eq:rrpos}
\end{equation}
These satisfy the recurrence relation
\begin{equation}
R_n=\frac{1}{f_l(n+\nu)-\kappa^2R_{n+1}}.
\end{equation}
In terms of these the $b_n$ can all be related to $b_0$ and expressed as
\begin{equation}
b_n=\left(\prod_{m=1}^n R_m\right)\kappa^n b_0.
\label{eq:bnpos}
\end{equation}
In a similar way, for negative $n$ I define 
\begin{equation}
{\overline R}_n=\frac{b_{-n}}{\kappa b_{-(n-1)}},
\end{equation}
and these satisfy
\begin{equation}
{\overline R}_n=\frac{1}{f_l(-n+\nu)-\kappa^2{\overline R}_{n+1}}.
\label{eq:rrneg}
\end{equation}
The corresponding coefficients can be written
\begin{equation}
b_{-n}=\left(\prod_{m=1}^n{\overline R}_m\right)\kappa^n b_0.
\label{eq:bnneg}
\end{equation}

Using these results in the remaining equation of Eq.~(\ref{eq:bdiffe}), 
with $n=0$, gives
\begin{equation}
f_l(\nu)-\kappa^2\left(R_1+{\overline R}_1\right)=0.
\label{eq:eigen}
\end{equation}
This is a nonlinear eigenvalue equation which determines the shift $\nu$.
Alternatively, if one is simply interested in the value of $\nu$, one can
look for the zeros of the infinite-dimensional Hill determinant of the
coefficients in Eq.~(\ref{eq:bdiffe}) (see Ref.~\cite{boamm}, Sec.~7.5).
If the set of equations is truncated to a finite number, this can be done
straightforwardly with the aid of Mathematica \cite{math}. Taking
$|n|\lesssim 20$ is sufficient to determine the zeros to six significant 
figures, at least for small angular momenta ($l\lesssim 5$).

The resulting eigenvalue equation, in either version, is an even function of
$\kappa$ and hence the roots of the equation are the same for both repulsive and 
attractive potentials of the same strength. From the fact that $f_l(\nu)$ is an 
odd function, it follows that 
\begin{equation}
{\overline R}_n(-\nu)=-R_n(\nu).
\label{eq:dsym}
\end{equation}
Hence if $\nu$ is a solution of Eq.~(\ref{eq:eigen}), so is $\nu$. 
Also, the roots are periodic under addition of any integer to $\nu$.
From now on, I shall use $\nu$ to denote the root whose real part lies 
between $l$ and $l+\frac{1}{2}$. This ensures that coefficient $a_0$ is large
in the expansion of the corresponding solution, $\phi^{(+)}(x)$, at least for
small $\kappa$. A second, independent solution, $\phi^{(-)}(x)$, is obtained 
by replacing $\nu$ by $-\nu$. From the symmetry of the $R_n$ it follows that
the coefficients in the two solutions are related by
\begin{equation}
a_{-n}^{(-)}=(-1)^{n} a_n^{(+)},
\end{equation}
if we choose
\begin{equation}
a_0^{(-)}=a_0^{(+)}\equiv a_0.
\end{equation}
The coefficient $a_0$ can be fixed by requiring that the solutions have the 
standard asymptotic normalisation of a Bessel function for large $r$.

Solutions to these equations can be obtained using perturbation theory
if $\kappa$ is small, corresponding to weak coupling or, equivalently, 
low energies. This is just the Born expansion, and it leads to expressions
for the solutions as power series in $\kappa$. The advantage of the Bessel
expansion \cite{cava,gao1,gao2} is that it determines the radius of convergence 
of this series, as well as providing analytic forms for the solutions in 
nonperturbative cases.

For $\kappa=0$ the order of the Bessel function is $\nu=l+\frac{1}{2}$, and 
the solutions of the radial Schr\"odinger equation are just spherical Bessel 
functions $j_l(pr)$. Consider next very small values of $\kappa$. For these 
we can approximate the ratios by their leading-order expressions from
Eqs.~(\ref{eq:rrpos},\ref{eq:rrneg}):
\begin{equation}
R_n\simeq\frac{1}{f_l(n+\nu)},\qquad {\overline R}_n\simeq\frac{1}{f_l(-n+\nu)},
\end{equation}
and the eigenvalue equation (\ref{eq:eigen}) becomes
\begin{equation}
f_l(\nu)-\kappa^2\left[\frac{1}{f_l(1+\nu)}+\frac{1}{f_l(-1+\nu)}\right]=0.
\end{equation}
Since the solution $\nu$ lies very close to $l+\frac{1}{2}$ in this limit, 
we can write
\begin{equation}
\nu=l+{\textstyle\frac{1}{2}}-\delta\nu.
\end{equation}
In partial waves with $l\ge 1$, we can approximate $f_l(n+\nu)$ by
\begin{eqnarray}
f_l(\nu)&\simeq& -4\left(l+{\textstyle\frac{1}{2}}\right)^2\delta\nu,\cr
\noalign{\vspace{5pt}}
f_l(n+\nu)&\simeq& 4n\left(l+n+{\textstyle\frac{1}{2}}\right)
\left(l+{\textstyle\frac{1}{2}}(n+1)\right).
\end{eqnarray}
Using these in the eigenvalue equation we get, at order $\kappa^2$,
\begin{equation}
\delta\nu=\frac{3}{16}\,\frac{\kappa^2}{\left(l-{\textstyle\frac{1}{2}}\right)
\left(l+{\textstyle\frac{1}{2}}\right)\left(l+{\textstyle\frac{3}{2}}\right)
l(l+1)}.
\end{equation}
In the corresponding solutions to the Schr\"odinger equation, only the coefficent 
$a_0$ is large. 

For $s$-waves we need to be more careful since $f_0(-1+\nu)$ is small in the
weak-coupling limit:
\begin{equation}
f_0(-1+\nu)\simeq -\delta\nu.
\end{equation}
Using this in the eigenvalue equation, we find that the leading shift $\delta\nu$ 
is of order $\kappa$, and is given by\footnote{Compare Ref.~\cite{sbcefmr} 
Sec.~2.5.2, noting that I have made a different choice from the multiple roots of 
the equation for $\nu$. For $s$-wave scattering by a pure $1/r^3$ potential, 
this dependence of the order of the Bessel functions on $\kappa$ leads to
nonanalytic terms in the phase shift \cite{ll,shake,gao2}. However these are not 
present for OPE, even in the chiral limit, since the tensor potential vanishes in 
$s$-waves.}
\begin{equation}
\nu={\textstyle\frac{1}{2}}-|\kappa|.
\end{equation}
In the corresponding wave function, $a_{-1}$ is of the same order in $\kappa$ 
as $a_0$. 

For larger values of $\kappa$ we need to iterate the recurrence relations
for the $R_n$ but, provided $\kappa$ is small enough, we can 
still expand the results as power series in $\kappa$. Similarly the solutions 
to the eigenvalue equation and hence the wave functions themselves can be found 
perturbatively as Born expansions in powers of $\kappa$ \cite{cava}. Since 
$\kappa \propto p B_3$, these expansions describe systems with either weak 
coupling or low energy. 

As $|\kappa|$ increases, these series converge more slowly and eventually
a perturbative treatment becomes impossible. A definite upper bound on the value
of $\kappa$ for which this occurs can be found by following the behaviour of the 
eigenvalues of Eq.~(\ref{eq:eigen}). As $|\kappa|$ increases, pairs of eigenvalues 
approach integer values from above and below until, at some critical value, they 
form degenerate pairs. Then, for larger values of $|\kappa|$, they move off into 
the complex plane \cite{gao2}. The presence of a square-root branch point where
this happens means that eigenvalue $\nu$ cannot be expanded in powers of $\kappa$
above this critical value, $\kappa_c$. If the eigenvalue cannot be expanded in 
this way then neither can the solutions to the set of linear equations 
(\ref{eq:bdiffe}), and hence $\kappa_c$ also provides an upper limit on 
the convergence of any perturbative expansion of the solutions. 

The values of $\kappa_c$ for low partial waves are listed in Table I. For large 
orbital angular momentum  ($l\gtrsim 20$) $\kappa_c$ grows roughly as $0.3\, l^3$. 
This can be understood if nonperturbative behaviour sets in when waves can penetrate
the centrifugal barrier to radii where the $1/r^3$ and centrifugal potentials
are roughly equal. To estimate where this occurs we can set 
\begin{equation}
E\simeq\frac{l(l+1)}{M_{\scriptscriptstyle N} r^2}\simeq\frac{B_3}{r^3}.
\label{eq:npcond}
\end{equation}
This leads to
\begin{equation}
\kappa\simeq[l(l+1)]^{3/2},
\end{equation}
which is consistent with the observed growth with $l$. 

\begin{table}[h]
\begin{tabular}{rlrrr}
\hline
\hbox{\quad} & $l$ &\hbox{\qquad} & $\kappa_c\quad$ & \hbox{\quad}\\
\hline
& 0 && 0.318058\\
& 1 && 2.51811\\
& 2 && 8.33342\\
& 3 && 19.6983\\
& 4 && 38.6026\\
& 5 && 67.0469\\
\hline
\end{tabular}
\caption{Critical values of of the dimensionless coupling $\kappa$ for which the 
eigenvalues $\nu$ form degenerate pairs. These agree with the values of 
$\epsilon_{sc}$ in Table I of \cite{gao2}, where $\epsilon_{sc}=\kappa_c^2/4$.}
\end{table}

This dependence of $\kappa_c$ on the orbital angular momentum demonstrates that 
it is short-distance physics that leads to the breakdown of perturbation theory:
the wave function has to penetrate the centrifugal barrier before it can ``see''
the singular core. The radius at which this happens is, from Eq.~(\ref{eq:npcond}),
of the order of
\begin{equation}
r_0\simeq\frac{\beta_3}{l(l+1)},
\end{equation}
where
\begin{equation}
\beta_3=M_{\scriptscriptstyle N}B_3
\end{equation}
is the length scale associated with the strength of the potential.
These estimates of the critical value of $\kappa$ and the associated radius
fail for small angular momenta. In particular $\kappa_c$ is nonzero for $l=0$. 
This is because the $1/r^3$ potential is short-ranged in the sense that, 
apart from a single logarithmic term, its low-energy scattering amplitude
can be expanded in powers of the energy \cite{gao2,ll,shake} 
(unlike the Coulomb or $1/r^2$ potentials). Hence a minimum momentum is 
required before its singular nature can be resolved.

In the realistic case of finite pion mass, the long-range tail of potential 
is replaced by an exponential fall-off. Since the breakdown of perturbation 
theory is a short-range effect, it should not be altered qualitatively by a 
nonzero $m_\pi$, especially in high partial waves. Provided the radius $r_0$ 
is much smaller than $1/m_\pi$, the finite mass will not even change the critical 
value of $\kappa$ significantly. In lower waves where $r_0$ comparable to 
or less than $1/m_\pi$, momenta of the order of $m_\pi$ are needed before 
the singularity is resolved and nonperturbative behaviour sets in.
Hence in channels where the momentum corresponding to $\kappa_c$
is less than $m_\pi$, such nonpertubative effects are still expected for 
momenta of order $m_\pi$, but not for much lower momenta.

The forms of the wave functions at short-distances will be an important 
ingredient in the RG analysis below and so I outline their basic
features here. More details of the solutions can be found in Ref.~\cite{gao2}, 
for the case of repulsive $1/r^3$ potentials. For small $r$, we can use the WKB
approximation to find their forms. In the case of a repulsive potential this
shows that the solutions have exponential dependence on $\sqrt{\kappa/x}$, or 
equivalently $\sqrt{\beta_3/r}$. In dimensionless form, the small-$x$ solutions 
are
\begin{equation}
\phi^{(\pm)}(x)\sim A^{(\pm)}(\kappa)\,x^{1/4}
\exp\left[2\sqrt{\frac{\kappa}{x}}\,\right]
+B^{(\pm)}(\kappa)\,x^{1/4}\exp\left[-2\sqrt{\frac{\kappa}{x}}\,\right].
\label{eq:asysmx}
\end{equation}

The determination of the coefficient $A$ of the dominant piece of
a solution requires careful asymptotic analysis of the series
Eq.~(\ref{eq:phiexp}). The important terms at small $x$ are those for
large negative values of $n$. These can be summed using Laplace's method (see
Ref.~\cite{boamm}, Sec. 6.7). Under the analytic continuation 
$x\rightarrow e^{i2\pi}x$ the dominant and subdominant pieces of the solution
exchange roles (an example of Stokes' phenomenon---see Ref.~\cite{boamm}, 
Sec.~3.7). This shows that their coefficients are related by
\begin{equation}
B^{(\pm)}=-ie^{\pm i2\pi\nu}A^{(\pm)}.
\end{equation}
Having found the small-$x$ forms of the two independent solutions 
$\phi^{(\pm)}(x)$, we can then build the regular solution (which behaves like 
$x^{1/4}\exp[-2\sqrt{\kappa/x}\,]$) as a linear combination of them.
These regular solutions form a complete, orthogonal set of basis functions for 
the RG analysis of the short-range physics.

In the case of an attractive inverse-cube potential, we need to be more careful.
Since, as already noted, $\nu$ and the ratios of determinants are independent 
of the sign of the potential, solutions for this case
can be obtained by replacing $\kappa$ by $-\kappa$ in the recursion relations 
for the ratios of coefficients $a_n$. Alternatively one can make an analytic 
continuation $x\rightarrow e^{i\pi}x$ of the solutions already found for the
repulsive case. The latter method shows that the solutions have the small-$x$
forms
\begin{equation}
\phi^{(\pm)}(x)\sim C^{(\pm)}(\kappa)\,x^{1/4}\cos\left[2\sqrt{\frac{\kappa}{x}}\,
+\left(\pm\nu-\frac{1}{4}\right)\pi\right],
\end{equation}
where I have now defined the dimensionless coupling to be positive-definite:
$\kappa= p M_{\scriptscriptstyle N}|B_3|$. The continuation also 
shows that the $C(\kappa)$ are related by
\begin{equation}
C^{(\pm)}=2A^{(\pm)},
\end{equation}
to the coefficients $A(\kappa)$ of the dominant pieces of solutions 
with the same $\nu$ for a repulsive potential of the same strength. 

Both of the solutions for an attractive potential 
display oscillatory behaviour as $x\rightarrow 0$ and 
so any linear combination of them is an equally good solution. For the RG analysis 
we need a well-defined set of orthogonal basis functions. As in the case of an 
attractive inverse-square potential,\footnote{See Refs.~\cite{bb2,bc} 
and references therein for more discussion of this potential.} this can
be obtained if we choose a self-adjoint extension of the original Hamiltonian. 
In practice this means fixing the phase of these short-distance oscillations 
\cite{pvra2,bbckmvk}. This phase should be independent of energy to form
an extension whose eigenfunctions are orthogonal. In essence one short-distance 
parameter, the leading, energy-independent term of the effective potential, 
has been used to provide a well-defined set of DW's of
the long-range potential. Clearly this leads to a redundancy in the 
parametrisation: a different choice of extension can be compensated by 
changing the leading term in the short-range potential \cite{bb2}. 
However any energy- or momentum-dependence due to short-range physics can be 
described entirely by higher-order terms in that potential.
Note that here I am requiring orthogonality simply in order to generate a
suitable DW basis for studying the effects of the short-range interactions.
This is in contrast to the approach developed in Refs.~\cite{pvra1,pvra2,pvra3}
where an orthogonality condition is imposed on the full wave functions and
this leads to very strong constraints on the short-distance interactions.

The forms of the short-distance wave functions for attractive and repulsive 
potentials look very different, depending sinusoidal or exponentially on 
$\sqrt{\beta_3/r}$. However we shall see that it is their power-law radial 
dependence that controls RG flow of short-distance interactions and this is same 
for both cases. Although the wave functions for the attractive potential oscillate
at short distances, these oscillations depend on a scale, $\beta_3$, unlike the 
analogous ones found for the inverse-square potential. 
The scale-free oscillations found 
there and in the corresponding three-body systems lead to limit cycles in the RG 
flow \cite{bb2,bbckmvk,bc,bp,bhvk,wilson,bh}, but here the scale-dependence of the 
$1/r^3$ potential means that we should not expect to find similar limit cycles.

\subsection{Coupled channels}

The solution of the Schr\"odinger equation for the coupled spin-triplet 
channels proceeds along very similar lines. Using the matrix elements of $S_{12}$
in the two-component basis of waves with $l=j\pm 1$ \cite{rs}, the chiral limit 
of the tensor potential can be written in the form
\begin{equation}
{\bf V}_{\pi T}(r)=\frac{1}{2j+1}\left(\begin{array}{cc}-2(j-1)&6\sqrt{j(j+1)}\cr 
6\sqrt{j(j+1)}&-2(j+2)\end{array}\right)\frac{B_T}{r^3},
\end{equation}
where $B_T=-3/(M_{\scriptscriptstyle N}\lambda_\pi)$ for isospin-singlet waves, 
with $l=j\pm 1$ even, and $B_T=1/(M_{\scriptscriptstyle N}\lambda_\pi)$ for 
isospin triplets, with $l=j\pm 1$ odd. Rescaling the equation as above, it 
can be written in a similar dimensionless 
form to Eq.~(\ref{eq:phiode}):
\begin{equation}
\left[x^2\,\frac{{\rm d}^2}{{\rm d}x^2}
+x\,\frac{{\rm d}}{{\rm d}x}+x^2\right]{\boldsymbol\phi}
-\left({\bf L}_j+{\textstyle\frac{1}{2}}{\bf 1}\right)^2{\boldsymbol\phi}
=\frac{1}{x}\,{\bf K}_j{\boldsymbol\phi},
\label{eq:phicodes}
\end{equation}
where the $2\times 2$ matrices are
\begin{equation}
{\bf L}_j=\left(\begin{array}{cc}j-1&0\cr 0&j+1\end{array}\right),\qquad
{\bf K}_j=\frac{\kappa_{\scriptscriptstyle T}}{2j+1}
\left(\begin{array}{cc}-2(j-1)&6\sqrt{j(j+1)}\cr 
6\sqrt{j(j+1)}&-2(j+2)\end{array}\right),
\label{eq:KL}
\end{equation}
and the dimensionless combination of momentum and coupling strength is 
\begin{eqnarray}
\kappa_{\scriptscriptstyle T}&=& p M_{\scriptscriptstyle N} 
B_{\scriptscriptstyle T}\cr
\noalign{\vspace{5pt}}
&=&\left\{\begin{array}{rl}-3p/\lambda_\pi&\quad\mbox{isospin singlet}\cr 
+p/\lambda_\pi&\quad\mbox{isospin triplet}\end{array}\right. .
\end{eqnarray}

The solutions to these equations can be expanded in Bessel functions as
\begin{equation}
{\boldsymbol\phi}(x)=\sum_{n=-\infty}^\infty {\bf a}_n J_{n+\nu}(x),
\end{equation}
where the ${\bf a}_n$ are two-component vectors. Substituting this into
Eq.(\ref{eq:phicodes}) leads to an infinite set of linear equations. As in
the uncoupled case, these can be put into a symmetric form by defining
${\bf b}_n={\bf a}_n/(n+\nu)$ and
\begin{equation}
{\bf F}(n+\nu)=\left(\begin{array}{cc}f_{j-1}(n+\nu)&0\cr
\noalign{\vspace{5pt}}
0&f_{j+1}(n+\nu)\end{array}\right).
\end{equation}
The resulting equations are then
\begin{equation}
{\bf K}{\bf b}_{n-1}-{\bf F}(n+\nu){\bf b}_n+{\bf K}{\bf b}_{n+1}=0\qquad 
-\infty<n<\infty.
\label{eq:bcdiffes}
\end{equation}

If we write
\begin{equation}
{\bf b}_n={\bf R}_n{\bf K}{\bf b}_{n-1},
\end{equation}
then the ${\bf R}_n$ satisfy the recurrence relation
\begin{equation}
{\bf R}_n=\left[{\bf F}(n+\nu)-{\bf K}{\bf R}_{n-1}{\bf K}\right]^{-1}.
\end{equation}
In terms of these matrices, the coefficients ${\bf b}_n$ for $n\ge 1$ are
\begin{equation} 
{\bf b}_n=\left(\prod_{m=1}^\infty {\bf R}_m{\bf K}
\right){\bf b}_0,
\end{equation}
where the matrix product should be read as starting with $m=1$ at the right.
Similarly we can write the coefficients ${\bf b}_{-n}$ as
\begin{equation} 
{\bf b}_{-n}=\left(\prod_{m=1}^\infty 
\overline{\bf R}_m{\bf K}\right){\bf b}_0,
\end{equation}
where the $\overline{\bf R}_n$ are given by
\begin{equation}
\overline{\bf R}_n=\left[{\bf F}(-n+\nu)-{\bf K}\overline{\bf R}_{n-1}
{\bf K}\right]^{-1}.
\end{equation}

Consistency of the equation for $n=0$ requires that $\nu$ satisfy
\begin{equation} 
\det\left[{\bf F}(\nu)-{\bf K}\Bigl({\bf R}_1+\overline{\bf R}_1\Bigr)
{\bf K}\right]=0.
\label{eq:eigen2}
\end{equation}
This is equivalent to the Hill determinant corresponding to Eq.~(\ref{eq:bcdiffes}). 
As in for the simple inverse-cube potential, the roots of this equation
are symmetric in $\nu$ and periodic. 

The $2\times 2$ nature of the problem means that this eigenvalue equation has two 
roots $\nu_1$ and $\nu_2$ with real parts between $l$ and $l+\frac{1}{2}$. The 
solutions to the Schr\"odinger equation (\ref{eq:phicodes}) for these roots will 
be denoted by ${\boldsymbol\phi}^{(1,+)}(x)$ and ${\boldsymbol\phi}^{(2,+)}(x)$, 
respectively. There are also two other independent solutions 
${\boldsymbol\phi}^{(1,-)}(x)$ and  ${\boldsymbol\phi}^{(2,-)}(x)$,
corresponding to the roots $-\nu_1$ and $-\nu_2$. The starting coefficients 
${\bf b}_0^{(i,\pm)}$ must be eigenvectors of the $n=0$ equation
\begin{equation} 
\left[{\bf F}-{\bf K}\Bigl({\bf R}_1 +\overline{\bf R}_1\Bigr){\bf K}
\right]_{\nu=\nu_i}{\bf b}_0^{(i,\pm)}=0.
\label{eq:n0eigen}
\end{equation}

Since ${\bf F}(n+\nu)$ is an odd function, the matrices for positive and 
negative roots are related by
\begin{equation}
\overline{\bf R}_n(-\nu)=-{\bf R}_n(\nu).
\end{equation}
If we choose the starting coefficients such that 
\begin{equation}
{\bf a}_0^{(i,-)}={\bf a}_0^{(i,+)}\equiv {\bf a}_0^{(i)},
\end{equation}
then the coefficients in the pairs of solutions are related by
\begin{equation}
{\bf a}_{-n}^{(i,-)}=(-1)^{n} {\bf a}_n^{(i,+)}.
\end{equation}
The ${\bf a}_0^{(i)}$ can again be fixed using the asymptotic normalisation
of the solutions.

As for the case of uncoupled channels, the eigenvalues $\nu$ move into the complex 
plane for large enough values of $\kappa_{\scriptscriptstyle T}$ and the solutions
can no longer be expanded perturbatively. The critical values 
for which this happens are listed in Table II. The coupled nature of the equations 
means that there are in general two of these values. The one exception is $j=0$, 
where only $l=1$ is possible and the critical value of $\kappa_{\scriptscriptstyle T}$ 
is just a quarter of the one in Table I for a $P$-wave.

\begin{table}[h]
\begin{tabular}{rlrrrrrrr}
\hline
\hbox{\quad} & $j$ &\hbox{\qquad} & $l$ &\hbox{\qquad} & $\kappa_{c1}\quad$ 
& \hbox{\quad}& $\kappa_{c2}\quad$ & \hbox{\quad}\\
\hline
& 0 && 1 && 0.629528\\
& 1 && 0,2 && 0.683495 && 2.48290\\
& 2 && 1,3 && 1.61857 && 6.91983\\
& 3 && 2,4 && 3.95647 && 14.3624\\
& 4 && 3,5 && 8.02206 && 23.2001\\
\hline
\end{tabular}
\caption{Critical values of of the dimensionless coupling 
$\kappa_{\scriptscriptstyle T}$ for which eigenvalues $\nu$ form degenerate pairs.}
\end{table}

The same methods outlined above can be used to find the coefficient vectors 
${\bf a}_n$ and hence to construct solutions to the radial Schr\"odinger 
equation (\ref{eq:phicodes}). Again, the full forms of these are not needed here, 
just their short-distance behaviours. These can be obtained by applying the WKB 
method to the eigenchannels of the potential. Since the eigenvalues of ${\bf K}_j$ 
are $+2\kappa_{\scriptscriptstyle T}$ and $-4\kappa_{\scriptscriptstyle T}$, one of 
these channels is repulsive and one attractive. For small $x$, all the solutions are
mixtures of exponential and sinusoidal pieces, with the form
\begin{eqnarray}
{\boldsymbol\phi}(x)&\sim& {\bf e}_+\,A(\kappa_T)\,x^{1/4}
\exp\left[2\sqrt{\frac{\kappa_+}{x}}\,\right]
+{\bf e}_+\,B(\kappa_T)\,x^{1/4}\exp\left[-2\sqrt{\frac{\kappa_+}{x}}\,\right]\cr
\noalign{\vspace{5pt}}
&&+\,{\bf e}_-\,C(\kappa_T)\, x^{1/4}\cos\left[2\sqrt{\frac{\kappa_-}{x}}\,
+\left(\pm\nu-\frac{1}{4}\right)\pi\right],
\end{eqnarray}
where $\kappa_+$ and $-\kappa_-$ are the eigenvalues of ${\bf K}_j$,
\begin{equation}
\kappa_+=\left\{\begin{array}{r}12p/\lambda_\pi\cr 
2p/\lambda_\pi\end{array}\right.,\qquad
\kappa_-=\left\{\begin{array}{rl}6p/\lambda_\pi&\quad\mbox{isospin singlet}\cr 
4p/\lambda_\pi&\quad\mbox{isospin triplet}\end{array}\right. ,
\label{eq:Kevalues}
\end{equation}
and ${\bf e}_\pm$ are the corresponding eigenvectors.
These eigenvectors are
\begin{equation}
{\bf e}_+=\frac{1}{\sqrt{2j+1}}\left(\begin{array}{c}\sqrt{j+1}\cr 
\sqrt{j}\end{array}\right),
\qquad {\bf e}_-=\frac{1}{\sqrt{2j+1}}\left(\begin{array}{c}\sqrt{j}\cr
-\sqrt{j+1}\end{array}\right),
\label{eq:Kevectors}
\end{equation}
for an isospin triplet channel, and \textit{vice versa} for a singlet.
As in the single-channel case, analytic continuation in $x$ can be used 
to relate $B(\kappa_T)$ to $A(\kappa_T)$, and to express $C(\kappa_T)$ in 
terms of the $A(\kappa_T)$ of the solution for a repulsive potential.

A basis set of physical, orthogonal solutions can formed by taking two linear
combinations of these four independent solutions at each energy. These must
be regular, with no admixture of the divergent $x^{1/4}\exp[2\sqrt{\kappa_+/x}]$ 
piece in the repulsive channel, as discussed in Refs.~\cite{pvra2,pvra3} for the 
$^3S_1$--$^3D_1$ waves. As for the attractive inverse-cube potential, one 
energy-independent parameter is also needed to fix the phase of the oscillations 
in the attractive channel \cite{bbsvk}. 
Although these solutions contain both oscillatory and exponentially decreasing 
pieces at small $x$, what matters in the RG analysis will be their power-law 
behaviour, which is the same as for the uncoupled $1/r^3$ case above.

\subsection{Critical momenta}

For the one-pion-exchange potential of interest here, the strengths of the tensor
interaction in the various channels are all fixed in terms of the scale $\lambda_\pi$ 
introduced in Eq.~(\ref{eq:lambdapi}). The critical values of the dimensionless
coupling can then be converted into critical values of the relative momentum for
each scattering channel. Taking $\lambda_\pi=290$~MeV leads to the critical
momenta listed in Table III. These are the maximum values of the momenta for which
one could attempt to construct a perturbative expansion of the solutions. In practice
one would expect such an expansion to be sufficiently convergent to be useful only
for momenta well below these values.

\begin{table}[h]
\begin{tabular}{rcrrr}
\hline
\hbox{\quad} & Channel &\hbox{\qquad} & $p_c\quad$ & \hbox{\quad}\\
\hline
& $^3S_1$--$^3D_1$ &&   66 MeV \\
& $^3P_0$          &&  182 MeV \\
& $^3P_1$          &&  365 MeV \\
& $^3P_2$--$^3F_2$ &&  470 MeV \\
& $^3D_2$          &&  403 MeV \\
& $^3D_3$--$^3G_3$ &&  382 MeV \\
& $^3F_3$          && 2860 MeV \\
& $^3F_4$--$^3H_4$ && 2330 MeV \\
& $^3G_4$          && 1870 MeV \\
\hline
\end{tabular}
\caption{Critical values of the relative momentum at which pairs of 
eigenvalues become degenerate and hence the tensor potential cannot be
treated perturbatively.}
\end{table}

From the table we see that in two channels, $^3S_1$--$^3D_1$ and $^3P_0$,
the critical values are less than or of the order of $m_\pi$. These are
obtained from the chiral limit of the OPE; the values for finite $m_\pi$ will
be somewhat higher. However they will still be of the order of $m_\pi$ since,
as discussed above, the exponential fall-off cannot make the radius at which 
the waves probe the singular core much smaller than $1/m_\pi$. It is thus
not surprising that Fleming, Mehen and Stewart \cite{fms} found that
the perturbative KSW approach fails for these cases. In the other $P$- and 
$D$-waves the values are low enough, $\sim 400$~MeV, to suggest that these 
should also be treated nonperturbatively at the energies of
interest for nuclear physics. Again this is in accord with the findings of
ref.~[9]. 

Between the $D$- and $F$-waves, the critical momenta jump by a factor of about 6.
Two effects contribute to this: the critical dimensionless couplings are about 
twice as large for the $F$-waves, and their physical couplings are three times 
smaller since the $F$-waves are isospin triplets. 
Hence for the $F$-waves and above that do not couple to
$P$- or $D$-waves, the breakdown scales are well above 1~GeV. In these cases 
it should be possible to treat OPE perturbatively. A similar conclusion is reached
in Ref.~\cite{ntvk}, but on the basis of very different arguments which rely 
heavily on keeping $m_\pi$ finite.

\section{Renormalisation-group analysis}

The long-range, pion-exchange physics in low-energy nuclear EFT's can be 
calculated from ChPT \cite{wein,border,bvkrev}. The short-range physics is then
parametrised in terms of contact interactions. If the EFT is to have any predictive 
power, we must be able to organise these interactions systematically according to 
some power counting. In weakly interacting systems one can do this by simply counting
powers of the low-energy scales, generically denoted here by $Q$. This naive 
dimensional analysis leads to the power counting used in
ChPT for mesons and single nucleons \cite{scherer}. In contrast, nonperturbative 
effects can introduce new low-energy scales and can generate anomalous dimensions 
for terms in the short-distance potential. As a result these terms may not scale 
as predicted by naive dimensional analysis and so the power counting can be quite 
different. 

The RG provides a general and powerful tool for analysing scale dependences,
particularly in nonperturbative systems. In the context of few-body systems, 
it is convenient to express this in the form of a differential equation describing 
the ``flow" of the short-range potential as the cut-off is varied 
\cite{bmr,bb1,bb2}. This equation can be constructed as follows.
First, we apply a floating cut-off at some scale $\Lambda$ that lies between the 
low-energy scales of interest and the scale of the underlying physics $\Lambda_0$. 
This assumes that these scales are well separated; if they are not, the expansion 
in powers of $Q/\Lambda_0$ will not converge and we shall not be able to construct 
a useful EFT. 
Second, we demand that physical observables be independent of the cut-off $\Lambda$,
since its value is arbitrary. As a result, the couplings in our effective potential 
must depend on $\Lambda$. This dependence ensures, in particular, that the couplings 
cancel any parts of loop integrals that diverge for large $\Lambda$. 
Finally, we rescale the theory by expressing all dimensioned quantities in units of
$\Lambda$. Powers of $\Lambda$ can then be used to determine the net powers of 
low-energy scales in the terms in the potential.

Having constructed the RG equation for the rescaled potential, we can look for fixed 
points. These are $\Lambda$-independent solutions that can form the end points of the
RG flow. They describe scale-free systems.
Perturbations around a fixed point can be expanded in eigenfunctions of the 
linearised RG equation that scale with definite powers of $\Lambda$, given by the 
eigenvalues of this equation. These perturbations can be classified according to 
their eigenvalues as relevant, irrelevant or marginal.

Marginal terms (``renormalisable" ones in field-theoretic language) have no power-law 
dependence on the cut-off after rescaling, although in general they can depend 
logarithmically on it. Like the fixed-point potential, these terms are important at 
all scales. They are the leading-order terms in the WvK scheme, of order $1/Q$ in 
low-energy scales. Since the loop integrals in the Lippmann-Schwinger equation are of 
order $Q$, all iterations of these terms are of the same order and hence they need 
to be treated nonperturbatively.
Irrelevant (or ``nonrenormalisable") terms vanish as positive powers of $\Lambda$
as $\Lambda\rightarrow 0$. These are higher-order terms which become weak at low 
energies and so can be treated perturbatively. 
Lastly, relevant (or ``super-renormalisable") terms grow as negative powers of
$\Lambda$. These are unimportant at high energies but become increasingly important 
at low energies, ultimately changing the nature of the low-energy EFT. If such 
terms are present, a fixed point is unstable and, for low-enough values of $\Lambda$, 
the theory will ultimately flow to a different point. An example of this is provided 
by the scattering-length term in the pionless EFT \cite{bmr}.

The RG approach developed in Ref.~\cite{bb1} assumes that two-body potential consists 
of a known long-range piece $V_L$ and a short-piece $V_S$ which parametrises the 
physics that lies outside the scope of our effective theory. This method starts
by using the ``two-potential trick" \cite{newton} to define a $T$-matrix describing 
scattering between distorted waves (DW's) of the long-range potential. The Hilbert 
space is then reduced by imposing a cut-off on the basis of DW's at momentum 
$\Lambda$. For this cut-off to lead a well-defined space, the DW's should form
a complete, orthogonal set of basis functions. This requires that the long-range
potential, and any parameter needed to form a self-adjoint extension of it,
should be independent of energy.

Demanding that the (fully off-shell) $T$-matrix be independent of the cut-off 
$\Lambda$ leads to the differential equation
\begin{equation}
\frac{\partial V_S}{\partial\Lambda}=-V_S\,\frac{\partial G_L}
{\partial\Lambda}\,V_S,
\label{eq:udwrge}
\end{equation}
where $G_L$ is the DW Green's function for the long-range potential.
In order to treat cases where the DW's vanish or diverge at the origin as
a result of nonperturbative effects of $V_L$, the short-range potential is
taken to have the $\delta$-shell form
\begin{equation}
V_S(p,\lambda,\Lambda,R;r)=V_S(p,\lambda,\Lambda,R)\,\frac{\delta(r-R)}{4\pi R^2},
\end{equation}
where $\lambda$ denotes a generic low-energy scale associated with $V_L$.
The radius $R$ provides a second regulator here, which can be thought of as a 
``factorisation scale" separating the long-range physics which lies within the 
domain of our EFT from the unknown nonperturbative physics at shorter distances.
This second scale is introduced because it leads to an RG equation from which the
scaling of the potential can be deduced in a particularly transparent way.
In approaches based on a simple momentum cut-off, as in Ref.~\cite{ntvk}, or a 
coordinate-space regulator, as in Refs.~\cite{pvra1,pvra2,pvra3}, a single scale 
plays both roles of renormalisation and factorisation.\footnote{This is analogous 
to the choice that is often made in studies of the QCD evolution of structure 
functions.} For present purposes, $R$ should be chosen to be small enough that 
the wave functions have reached a common, energy-independent form.

With this $\delta$-shell form for $V_S$, Eq.~(\ref{eq:udwrge}) becomes
\begin{equation}
\frac{\partial V_S}{\partial\Lambda}=-\,\frac{M_{\scriptscriptstyle N}}{2\pi^2}\,
|\psi_L(\Lambda,R)|^2\,\frac{\Lambda^2}{p^2-\Lambda^2}\,
V_S^2(p,\lambda,\Lambda,R),
\label{eq:udwrge2}
\end{equation}
where $\psi_L(p,r)$ are the DW's for $V_L$ and I have assumed that it does not 
produce any bound states.
Taking the ``factorisation" radius $R$ to be small enough that it lies in the 
asymptotic region, the DW's can all be written in the form
\begin{equation}
\psi_L(p,R)\sim\mathcal{N}(\lambda/p)(pR)^{(\sigma-1)/2}F(\lambda R).
\label{eq:wfform}
\end{equation} 
This is a slight generalisation of the cases considered in Ref.~\cite{bb1}
to potentials that generate a non-power-law dependence on $R$ in this region,
described by the function $F(\lambda R)$.

Following the general method of Ref.~\cite{bb1}, I introduce a rescaled on-shell
momentum, $\hat p=p/\Lambda$. Other low-energy variables are treated similarly, 
and I then define the rescaled potential,
\begin{equation}\label{eq:dwscale}
\hat V_S(\hat p,\hat\lambda,\Lambda)=\frac{M_{\scriptscriptstyle N} \Lambda}
{2 \pi^2}\,(\Lambda R)^{\sigma-1}\,|F(\Lambda\hat\lambda R)|^2\,
V_S(\Lambda\hat p,\Lambda\hat\lambda,\Lambda,R).
\end{equation}
Inserting this and Eq.~(\ref{eq:wfform}) into the differential equation
Eq.~(\ref{eq:udwrge2}) gives the RG equation for $\hat V_S$:
\begin{equation}
\Lambda\,\frac{\partial\hat V_S}{\partial\Lambda}
=\hat p\,\frac{\partial\hat V_S}{\partial\hat p}
+\hat\lambda\,\frac{\partial\hat V_S}{\partial\hat\lambda}
+\sigma\hat V_S
+\frac{|\mathcal{N}(\hat\lambda)|^2}{1-\hat p^2}\,\hat V^2_S.
\label{eq:dwrge}
\end{equation}
Since the function $|F(\Lambda\hat\lambda R)|^2$ depends on the product
$\Lambda\hat\lambda$, its derivatives cancel in this equation. The evolution of
$\hat V_S$ thus depends only on the power-law part of the DW's.

Note that in Eq.~(\ref{eq:dwscale}) I have implicitly demanded $V_S$ depend
on $R$ in such a way that the rescaled potential be independent of $R$. 
This implies that $|\psi_L(p,R)|^2V_S$ is also independent of $R$, and hence 
it ensures that scattering observables do not depend on this arbitrary radius. 
This is very similar to the philosophy adopted by Pav\'on Valderrama and Ruiz
Arriola in Refs.~\cite{pvra1,pvra2,pvra3}, 
except that there the coordinate-space regulator is the only one. However
the complicated dependence of $F(\lambda R)$ on $R$ means that the scaling
behaviour of the terms in the potential cannot easily be determined from
this condition. In the present approach the radial regulator separates 
off the regime of nonperturbative short-distance physics, while the momentum
cut-off $\Lambda$ is used to analyse the scale dependence.
For this to work, $R$ must be in the region where the DW's $ \psi_L(p,R)$ have
reached their energy-independent asymptotic form, at least for $p\leq\Lambda$.

The starting points for analysing the scale dependence of the potential
are the fixed points of the RG, solutions to eq.~(\ref{eq:dwrge}) that are 
independent of $\Lambda$. Expanding $\hat V_S$ around one of these points
and keeping only linear terms in RG equation, we get an equation whose
eigenfunctions scale with definite powers of $\Lambda$. Since the rescaling
means that the power of $\Lambda$ counts the net power of all low-energy scales
in each term, these eigenvalues determine the power counting for the terms in the
potential. For example, in the case of a pure short-range potential \cite{bmr},
the power counting for the expansion around the trivial fixed point is just the 
one originally proposed by Weinberg\cite{wein}. There is also a nontrivial fixed
point, which describes a system with an infinite scattering length. The expansion
around this point can be organised according to the power counting developed
in Refs.~\cite{ksw,bvk,vk}. The terms in this are in one-to-one correspondence
with the terms in the effective-range expansion \cite{bethe,newton}.

The form of the RG equation above can easily be used to analyse scale dependences in 
the vicinity of the trivial fixed point, $\hat V_S=0$. To find and study nontrivial 
fixed points, it is more convenient to convert it into a linear equation for 
$1/\hat V_S$ \cite{bb1}:
\begin{equation}
\Lambda\,\frac{\partial}{\partial\Lambda}\left(\frac{1}{\hat V_S}\right)
=\hat p\,\frac{\partial}{\partial\hat p}\left(\frac{1}{\hat V_S}\right)
+\hat\lambda\,\frac{\partial}{\partial\hat\lambda}
\left(\frac{1}{\hat V_S}\right)-\sigma\,\frac{1}{\hat V_S}
-\frac{|{\cal N}(\hat\lambda)|^2}{1-\hat p^2}.
\label{eq:dwrge3}
\end{equation}
The detailed forms of the fixed-point solutions to this equation are not necessary 
to determine the power countings for perturbations around them. If needed, they can 
be found by applying the methods of Refs.~\cite{bb1,bb2,barth}. These rely on the 
fact that the basic loop integral from the Lippmann-Schwinger equation satisfies 
the $\Lambda$-independent version of Eq.~(\ref{eq:dwrge3}). This integral
contains a piece that is a nonanalytic function of $\hat p/\hat\lambda$. 
Cancelling off this piece then leaves a well-behaved solution, $\hat V_{S0}$, 
which is analytic in the low-energy variables, $\hat p$ and $\hat\lambda$. 

So far I have discussed only energy-dependent perturbations. More generally, 
as an off-shell quantity, the short-range effective potential can also depend 
on momenta. This is equivalent to including terms with spatial derivatives in 
$V_S$. For the expansion of a pure short-range potential around the trivial fixed
point, the power counting is the same for both energy- and momentum-dependent
terms. Hence one can ``use the equation of motion", to exchange one dependence 
for the other \cite{fear}.

The expansion around around a nontrivial fixed point is more complicated.
There the equation of motion involves the fixed-point potential (which can 
include both long- and short-range pieces) and so purely momentum-dependent
terms are not eigenfunctions of the linearised RG equation, in contrast to the
energy-dependent ones discussed above. The coefficients of the latter terms 
are directly related to on-shell scattering observables through an 
effective-range expansion, as discussed in Refs.~\cite{bmr,bb1}. There are also 
terms with mixed momentum- and energy-dependence that scale with definite powers 
of $\Lambda$. However these affect only the off-shell behaviour of the scattering 
amplitude. Furthermore, they are of higher order than the corresponding 
energy-dependent terms \cite{bmr}. Hence, if one expands a momentum-dependent
term in eigenfunctions of the RG, the scaling with $\Lambda$ of the dominant
piece is governed by the eigenvalue of the corresponding energy-dependent term.
This is why the analyses of Refs.~\cite{bvk,vk,ksw}, using momentum-dependent
potentials, arrive at the same power counting as Ref.~\cite{bmr}, which uses
energy-dependent ones.

\subsection{Central OPE}

The first step in constructing any EFT is to identify all the important
low-momentum scales. To illustrate the choices involved in the nucleon-nucleon 
system, I consider first the central piece of the OPE potential and summarise the 
relevant results from Ref.~\cite{bb1}. For scattering at energies $\sim 100$~MeV 
the relative momentum and the pion mass obviously form two of these scales. For the 
central Yukawa potential, we can construct the scale $\alpha_\pi=m_\pi^2/\lambda_\pi
\simeq 70$~MeV. In strict chiral power counting this would be of order $Q^2$,
since it contains two powers of $m_\pi$. However if we choose to treat $\lambda_\pi$ 
as an additional low-energy scale, then $\alpha_\pi$ is promoted to order $Q$.

To see the consequences of this in the context of the RG for the $^1S_0$ channel, 
we can multiply the effective Hamiltonian by $M_{\scriptscriptstyle N}$ and define 
the dimensionless potential
\begin{equation}
\hat V_{\pi {\scriptscriptstyle C}}(r)=\frac{M_{\scriptscriptstyle N}
V_{\pi {\scriptscriptstyle C}}(r)}{\Lambda^2}.
\end{equation}
Then we need to express all low-energy scales in units of $\Lambda$.
If we regard $\lambda_\pi$ as a high-energy scale, then we define a
rescaled on-shell momentum by $\hat p=p/\Lambda$, and similarly 
$\hat m_\pi=m_\pi/\Lambda$ and $\hat r=\Lambda r$.\footnote{This is just the 
coordinate-space version of the rescaling discussed above.}
The resulting rescaled potential,
\begin{equation}
\hat V_{\pi {\scriptscriptstyle C}}(\hat r)
=-\Lambda\,\frac{\hat m_\pi^2}{\lambda_\pi}
\,\frac{e^{-\hat m_\pi \hat r}}{\hat r},
\end{equation}
is of order $\Lambda$, and hence is an irrelevant perturbation in the RG sense
that it vanishes as $\Lambda\rightarrow 0$. This is the choice made in the KSW 
scheme. In contrast, if we treat $\lambda_\pi$ as a low-energy scale and express 
it in units of $\Lambda$, writing $\lambda_\pi=\Lambda\hat\lambda_\pi$, then the 
rescaled potential is independent of $\Lambda$ and so forms part of any fixed 
point of the RG. In Weinberg's power counting \cite{wein}, this choice means that 
the potential (in momentum space) is of order $1/Q$.\footnote{An
alternative way to arrive at this result is to assign the nucleon mass an 
order $1/Q$ in the power counting \cite{wein,epel}. This also leads to $\lambda_\pi$
being identified as a low-energy scale of order $Q$.}
It must thus be iterated to all orders when solving the 
Schr\"odinger equation, along with the leading contact interaction, and so this 
choice corresponds to the WvK scheme. 

In the $^1S_0$ channel, the DW's of the Yukawa potential tend to constants
as $R\rightarrow 0$ and so have the short-distance form
\begin{equation}
\psi_L(p,R)\sim {\mathcal{N}}(\alpha_\pi/p, m_\pi/p).
\end{equation}
This corresponds to setting $\sigma=1$ and $F(\lambda R)=1$ in Eq.~(\ref{eq:wfform}) 
above. The resulting RG equation for $\hat V_S$ is given by Eq.~(\ref{eq:dwrge}), 
with $\sigma=1$ and three low-energy scales, $\hat p$, $\hat\alpha_\pi$ and 
$\hat m_\pi$. 

The expansion of the potential around the trivial fixed point has the form
\begin{equation}
\hat V_S=\sum_{k,m,n} C_{kmn}\Lambda^\rho\,
\hat m_\pi^{2k}\, \hat \alpha_\pi^m\, \hat p^{2n}.
\end{equation}
The RG eigenvalues of these terms are $\rho=2k+m+2n+1$ where $k$, $m$ and $n$ 
are non-negative integers. The corresponding power counting assigns them
orders $Q^d$ where $d=\rho-1$ \cite{bmr,bb1}. As in the case of a pure short-range 
potential, their eigenvalues start at $\rho=1$ and so they are all irrelevant.
This would be the appropriate power-counting if all short-range interactions in the 
$^1S_0$ channel were weak.

In the RG framework, the WvK treatment of this channel corresponds to an 
expansion of the short-distance potential around the nontrivial fixed 
point $\hat V_{S0}$. This can be written in the form
\begin{equation}
\frac{1}{\hat V_S}=\frac{1}{\hat V_{S0}}-\sum_{k,m,n} C_{kmn}\Lambda^\rho\,
\hat m_\pi^{2k}\, \hat \alpha_\pi^m\, \hat p^{2n}.
\end{equation}
The RG eigenvalues of these terms are $\rho=2k+m+2n-1$ where $k$, $m$ and $n$ 
are non-negative integers. This is similar to the expansion around 
the nontrivial fixed point for a pure short-range potential \cite{ksw,bmr} in
that the leading perturbation is a relevant one, with eigenvalue $\rho=-1$.
The terms are in one-to-one correspondence with the terms in a DW or ``modified" 
effective-range expansion \cite{bethe,vhk,kr,sf}. In it, all rapid energy dependence 
associated with the low-energy scales of OPE is factored out, to leave an 
amplitude whose energy dependence is controlled only by scales from the short-range 
physics.

In this channel we therefore have a choice between the two schemes, both
of which lead to consistent expansions of the low-energy physics. The KSW scheme
suffers from poor convergence because of the smallness of $\lambda_\pi$, while 
the WvK one, by treating $\lambda_\pi$ as a low-energy scale, converges better
but lacks a clear connection with ChPT. The most immediate signal of the latter 
problem is the contact interaction needed to renormalise the logarithmic 
divergence produced by the $1/r$ singularity of the potential. This is proportional
to $\alpha_\pi$, which contains two powers of $m_\pi$ in the 
chiral expansion but which must be treated as a single small scale in the
the WvK scheme \cite{bb1}. This is discussed by Beane {\it et al.}~\cite{bbsvk}, 
who conclude that the KSW scheme should be used in the $^1S_0$ channel. However, it 
is worth stressing that the problem is a lack of consistency with the chiral
power counting for other effective operators, not an internal inconsistency of
the DW effective-range expansion.

\subsection{Tensor OPE}

The nonperturbative nature of the tensor piece of OPE at short distances means 
that the RG analysis is rather different from that for the central piece just
discussed. In Sec.~II we have seen that the wave functions in all partial waves 
(and for both attractive and repulsive potentials) have the same power-law 
dependence on $x=pr$ at short distances. If we choose $R$ to be small enough 
that it lies in the asymptotic WKB region, then the uncoupled DW's can all be 
written in the form
\begin{equation}
\psi_L(p,R)\sim\mathcal{N}(\lambda_\pi/p)\,(pR)^{-1/4}F(\lambda_\pi R),
\end{equation} 
where $F(\lambda_\pi R)$ is the non-power-law part of the wave function.
In the isospin-triplet waves, where the tensor potential is repulsive, 
this function is
\begin{equation}
F(\lambda_\pi R)=\exp\left[-2\sqrt{\frac{2}{\lambda_\pi R}}\,\right],
\end{equation}
In the isospin-singlet channels, we have
\begin{equation}
F(\lambda_\pi R)=\cos\left[2\sqrt{\frac{6}{\lambda_\pi R}}
+\gamma\right],
\end{equation}
where $\gamma$ is an energy-independent phase. Here, for simplicity, 
I will continue to consider the chiral limit of the tensor OPE 
potential. Allowing for a finite $m_\pi$ will not alter the form of 
the short-distance wave functions, but it will introduce a dependence on
$m_\pi/p$ into the normalisation constants $\mathcal{N}$.

For the coupled channels, we have two independent solutions which satisfy the 
two boundary conditions as needed to generate an orthognal set of DW's, namely
that as $r\rightarrow 0$ the exponential piece should be regular and 
the oscillatory piece should have an energy-independent phase $\gamma$.
These solutions can be written
\begin{equation}
{\boldsymbol\psi}_{Li}(p,R)\sim \mathcal{N}_{i+}(\lambda_\pi/p)\,(pR)^{-1/4}
F_+(\lambda_\pi R)\,{\bf e}_+
+\mathcal{N}_{i-}(\lambda_\pi/p)\,(pR)^{-1/4}
F_-(\lambda_\pi R)\,{\bf e}_-,\qquad i=1,2,
\end{equation} 
where the short-distance forms of the wave functions are
\begin{equation}
F_+(\lambda_\pi R)=\exp\left[-2{\displaystyle\sqrt{\frac{\kappa_+}{x}}}\,\right],
\qquad F_-(\lambda_\pi R)=\cos\left[2{\displaystyle\sqrt{\frac{\kappa_-}{x}}}
+\gamma\right],
\end{equation}
and $\kappa_\pm$ and ${\bf e}_\pm$ are the eigenvalues and eigenvectors of the 
potential matrix ${\bf K}_j$, defined in 
Eqs.~(\ref{eq:Kevalues},\ref{eq:Kevectors}).

Matrix elements of the short-range potential between DW's can be written
\begin{equation}
\langle\psi_{Li}|V_S|\psi_{Lk}\rangle
=\sum_{\alpha,\beta=\pm}\mathcal{N}_{i\alpha}(\lambda_\pi/p)^*
F_\alpha(\lambda_\pi R)^*\,V_{S\alpha\beta}(p,\lambda_\pi,\Lambda,R)
F_\beta(\lambda_\pi R)\mathcal{N}_{k\beta}(\lambda_\pi/p),
\end{equation}
where the strength has been expressed as a matrix
${\bf V}_S(p,\lambda_\pi,\Lambda,R)$ using the basis of the eigenvectors 
${\bf e}_\pm$.

By analogy with the single-channel case, I define a rescaled potential 
$\hat{\bf V}_S$ as a matrix with elements
\begin{equation}
\hat V_{S\alpha\beta}=\frac{M_{\scriptscriptstyle N} \Lambda}
{2 \pi^2}\,(\Lambda R)^{-1/2}\,F_\alpha(\lambda_\pi R)^*\,
V_{S\alpha\beta}(p,\lambda_\pi,\Lambda,R)F_\beta(\lambda_\pi R).
\end{equation}
It is also convenient to define the matrix ${\bf N}(\lambda_\pi/p)$
with elements
\begin{equation}
N_{\alpha i}=\mathcal{N}_{i\alpha}(\lambda_\pi/p).
\end{equation}
Using this in the two-component version of Eq.~(\ref{eq:udwrge2}) leads to the
RG equation for $\hat{\bf V}_S$,
\begin{equation}
\Lambda\,\frac{\partial\hat {\bf V}_S}{\partial\Lambda}
=\hat p\,\frac{\partial\hat {\bf V}_S}{\partial\hat p}
+\hat\lambda_\pi\,\frac{\partial\hat {\bf V}_S}{\partial\hat\lambda_\pi}
+\frac{1}{2}\,\hat {\bf V}_S
+\frac{1}{1-\hat p^2}\,\hat {\bf V}_S\, {\bf N}(\hat\lambda_\pi)
{\bf N}(\hat\lambda_\pi)^\dagger\,\hat {\bf V}_S.
\label{eq:cdwrge}
\end{equation}

Apart from the additional matrix structure, this equation has the same form as 
the RG equation (\ref{eq:dwrge}) that governs the potential in the 
uncoupled channels. In particular, the coefficient $\sigma$, which arises from
the power-law behaviour of the wave functions at short distances, is the same 
in all cases (attractive, repulsive and coupled). This means that the scaling
behaviours of perturbations are the same in all cases, although obviously the 
forms of any nontrivial fixed points are different. In writing down these equations,
I have ignored the possibility of bound states in the attractive channels. 
I discuss below the modifications needed to take account of these, but they do not 
affect the scaling behaviour.

As always, each of these RG equations has a trivial fixed-point solution 
$\hat V_S=0$. The expansion around this has the form
\begin{equation}
\hat V_S=\sum_{m,n}C_{mn}\Lambda^\rho\,\hat\lambda_\pi^m\,\hat p^{2n},
\end{equation}
where $n$ and $m$ are non-negative integers, and the RG eigenvalues of the terms are
$\rho=m+2n+\frac{1}{2}$. They are all irrelevant perturbations, in the sense that 
their eigenvalues are positive, and so the fixed point is stable. The appearance of 
noninteger anomalous dimensions should not be surprising in the context of the RG. 
The version developed in Ref.~\cite{bb1} shows that the scaling is controlled by the 
power-law behaviour of the wave functions at short distances and in general this is 
noninteger for potentials with an inverse-power-law form.\footnote{Indeed other 
examples of noninteger anomalous dimensions have recently been found in the context 
of three-body systems \cite{grie}.} In the present case, half-integer values
appear because $|\psi|^2\sim(pR)^{-1/2}$.

For comparison, the terms in a pure short-range potential describing weak scattering 
in an $S$-wave have RG eigenvalues $\rho=2n+1$ \cite{bmr}. From this we see that the 
corresponding terms in the presence of a $1/r^3$ potential have eigenvalues
that are smaller by subtraction of $1/2$ and so they vanish more slowly as 
$\Lambda\rightarrow 0$. This means that the effects of these interactions have been 
enhanced by the $1/r^3$ potential. (In the language of the RG, they are more 
``relevant".) If we translate these results into the more usual power counting, 
we find that these terms have orders $Q^{m+2n-1/2}$, and so they have been promoted
by half an order compared with the power counting one would get from naive
dimensional analysis ($Q^{m+2n}$). 

In the $S$-, $P$- and $D$-waves, the radii at which the OPE and centrifugal 
potentials become comparable are of the order of 1~fm. 
By choosing the factorisation radius $R$ to lie significantly inside this, 
we can ensure that the short-range potential acts in a region where the wave 
functions have reached their asymptotic, WKB form for the $1/r^3$ potential. 
In the $^3P_1$ and $^3D_2$ channels, for example, this occurs for radii of the 
order of 0.1~fm or less. At these distances the wave function in the
$^3P_1$ channel is is highly suppressed by the repulsive $1/r^3$ potential.
Somewhat larger radii, of the order of 0.4~fm, can also be used since the 
wave functions still have energy-independent forms,\footnote{By neglecting 
the $x^2$ term on the LHS and changing variables to $y=\sqrt{\kappa/x}$, 
Eq.~(\ref{eq:phiode}) can be put into the form of Bessel's equation. This 
shows that the energy-independent solutions can be expressed in terms of 
order-$2l+1$ Bessel functions of $y$. For large $y$, these tend to the asymptotic 
WKB expressions discussed above.} at least for energies 
up to about 250~MeV, but even here the $^3P_1$ wave function is already
small, down by a factor of 5--10 compared to its size in the region 
0.7--1~fm. 

The corresponding short-range potential must therefore be enhanced by a large 
numerical factor if we choose $R$ less than 0.4~fm. However one should remember 
that the potential is not a physical quantity. In any observable, $V_S$ always 
appears multiplied by two short-distance wave functions. Indeed the radius $R$
is arbitrary and the form of the $R$-dependence of $V_S$ was chosen to ensure 
that scattering observables are independent of it. It is the size of the physical 
effects of a term in $V_S$ that matters and, in particular, whether these require 
that it be iterated to all orders, or allow it to be treated perturbatively. 
Such questions are answered by the power counting that the RG analysis provides.

Note that, despite also 
using a radial cutoff, the approach here is quite different from that Pav\'on
Valderrama and Ruiz Arriola \cite{pvra1,pvra2,pvra3}: those authors consider 
the limit as their radial cutoff tends to zero and so they set all irrelevant 
perturbations to zero. As a result, their predictions for scattering observables 
just depend on the long-range pion-exchange potential and a small number of 
short-distance parameters associated with any relevant or marginal terms.

The results above apply to all partial waves where tensor OPE is treated 
nonperturbatively. Hence in waves with nonzero orbital angular momentum, where 
the leading terms are naively of order $Q^{2l}$, the orders of these terms are 
much lower than dimensional analysis would suggest. This agrees with the conclusion 
of Nogga, Timmermans and van Kolck \cite{ntvk} that, based on their numerical 
analysis, short-range terms must be promoted in channels where the tensor potential 
is attractive. The RG analysis here makes quantitative the degree of promotion 
involved by determining the power counting for all terms in the double expansion in 
powers of energy ($p^2$) and the coupling scale ($\lambda_\pi$). It also shows that 
the effect is present in repulsive as well as attractive channels.

The scattering in the $^3S_1$--$^3D_1$ channel is strong at low energies and 
so the trivial fixed point is not an appropriate starting point. Instead we need
to find a nontrivial fixed point and expand around it. This is most easily 
done by rewriting the RG equation as in Eq.~(\ref{eq:dwrge3}).
The linear nature of this equation makes it straightforward to find
the perturbations around the fixed point that scale with definite
powers of $\Lambda$. The resulting expansion is
\begin{equation}
\frac{1}{\hat V_S}=\frac{1}{\hat V_{S0}}-\sum_{m,n}C_{mn}
\Lambda^{m+2n-1/2}\,\hat\lambda_\pi^m\,\hat p^{2n},
\label{eq:ntpert}
\end{equation}
where $n$ and $m$ are again non-negative integers. As in the case of a pure 
short-range potential \cite{bmr}, the nontrivial fixed point is unstable, with one 
negative eigenvalue. Terms in this expansion can be related to the terms in a DW 
effective-range expansion, analogously to the examples studied in 
Refs.~\cite{bb1,bb2}. The RG eigenvalue for a general term in Eq.~(\ref{eq:ntpert})
is $\rho=m+2n-1/2$, which should be compared with $\rho=2n-1$ for the pure
short-range case \cite{bmr}. This shows that the terms in the expansion here have 
been demoted by half an order (that is, they are less important) compared with the 
corresponding terms without the long-range potential. 

Although strictly the RG equation (\ref{eq:dwrge}) only applies to channels
where the tensor force is repulsive, the scaling behaviour is in fact the same 
for the attractive and coupled channels. The only difference is that
that attractive $1/r^3$ potentials give rise to deeply bound states that lie
outside the domain of our EFT. We should therefore cut them off at 
$E=-\Lambda^2/M_{\scriptscriptstyle N}$, as in the case of the attractive 
inverse square potential \cite{bb2}. This adds $\delta$-function terms 
to the RG equation at the values of $\Lambda$ where bound states fall outside 
the cut-off. These lead to step discontinuities in $\hat V_S$ at these points, 
which can be thought of as jumps to different branches of the fixed-point
potential $\hat V_{S0}$. 

The existence of multiple branches of the potential is a consequence of the 
oscillatory nature of the short-distance wave functions. In order to make
these well defined, we had to choose a particular self-adjoint extension of the 
long-distance Hamiltonian by fixing the phase of these oscillations. As in the 
$1/r^2$ case, the energy-independent short-range interaction has the effect of 
changing the self-adjoint extension \cite{bb2}. However, in that example, the 
scale-free nature of the potential and the associated Efimov effect \cite{efimov}
(an infinite tower of geometrically spaced bound states) mean that the
different choices lie on a limit cycle of the RG. As a result the leading  
short-range term forms a marginal perturbation which changes the 
starting point on that cycle. In contrast, only a discrete set of extensions 
of the $1/r^3$ Hamiltonian lead to scale-free systems with bound states at
zero energy. The different branches of $\hat V_{S0}$ correspond to this set of
extensions.

The use of the DW basis in this RG analysis makes it straightforward to expand
the potential in terms of perturbations that scale with definite powers 
of $\Lambda$. In Ref.~\cite{ntvk} the cut-off was applied to a plane-wave basis. 
Such a cut-off has two effects: it both regulates the short-distance interaction, 
and it removes the singularity of the long-range potential at the origin. The 
second aspect means that changing the cut-off has the effect of changing the 
self-adjoint extension that determines the long-range behaviour of the DW's. 
After fitting to low-energy scattering observables, the resulting short-distance 
potential displays dramatic oscillations due to the changing number of bound 
states of the long-range potential. Similar oscillations are also seen with 
the radial cut-off of Ref.~\cite{pvra2}. However, with care, it should still be 
possible to determine the power counting using such cut-offs, as illustrated by the
analysis of attractive three-body systems in Ref.~\cite{bghr}.

In spin-triplet channels without low-energy bound or virtual states (in other 
words, all except $^3S_1$--$^3D_1$), we can define the DW basis by picking an 
initial extension that does not produce a low-energy bound or virtual state and 
then we can expand the short-range potential around the trivial fixed point. The
terms in this expansion are all irrelevant. This implies that, provided we pick an 
initial extension that gives weak low-energy scattering, any dependence of the 
low-energy phase shifts on this choice must be small. Otherwise the leading
irrelevant perturbation could not be equivalent to a change in the extension.

This weak dependence of scattering observables on the choice of extension, at least 
well away from the ones that generate low-energy bound states, can be seen in the 
numerical results of Ref.~\cite{ntvk}, particularly in Figs.~10 and 12. These 
contain long ``plateau regions" where the short-distance interactions are small and
only weakly dependent on $\Lambda$. These values of the cut-off correspond to
extensions that generate only deeply bound states, outside the scope of the EFT.
In contrast there are also narrow ranges of $\Lambda$ where the regulated tensor
potential produces a low-energy bound or virtual state. To describe the observed
weak scattering, the potential in these regions must be supplemented by the
nontrivial fixed point and its relevant (unstable) perturbation, hence the very
large counterterms needed.

The results so far can be applied to the $S$-, $P$- and $D$-waves where, 
as we saw in Sec.~II, the tensor OPE needs to be treated nonperturbatively. 
As discussed above, the wave functions in these channels attain their
energy-independent short-distance forms for radii of 0.4~fm or less. These
forms are controlled by the $1/r^3$ tensor potential which is stronger than 
the centrifugal barrier for radii smaller than about 1~fm.
In $F$-waves and above, in contrast, the OPE and centrifugal potentials become 
comparable at radii of the order of 0.1--0.2~fm or less. The asymptotic forms
of the waves controlled by the tensor potential are thus reached only for radii
much less than 0.1~fm, far beyond the domain of validity of our EFT.
Moreover, the amplitudes of the wave functions for such small radii will be 
strongly suppressed by the centrifugal barrier, at least for momenta of the order 
of $m_\pi$.\footnote{The only exception would be if we were to choose an extension 
of the attractive $1/r^3$ potential that leads to a low-energy bound or virtual 
state which is trapped inside the barrier. However there is no reason to do so
in the context of nucleon-nucleon scattering in high partial waves.} In effect 
the strong centrifugal barrier in high partial waves ``protects" low-energy waves 
from probing the nonperturbative region. 

We can take advantage of this by choosing our factorisation scale $R$ for the 
high partial waves to be of the order of 0.4~fm, so that it lies in the region 
where the centrifugal potential dominates over both the tensor OPE and the 
on-shell energy. This allows us to treat OPE as a perturbation and 
to expand the DW's in Eq.~(\ref{eq:udwrge2}) in powers of $V_{\pi T}$. 
In this region the waves still have the $r^l$ form produced by the 
centrifugal barrier and so scaling behaviour of the short-range potential
is the same as that in the presence of the $1/r^2$ centrifugal
potential alone \cite{bb1}. In particular the resulting power counting is 
just the usual one given by naive dimensional analysis.
Of course, at high enough energies, the waves will penetrate the barrier
and this perturbative treatment will break down. However, in the high partial 
waves, the results in Sec.~IIC show that this only happens at scales that 
lie outside the domain of our low-energy effective theory. 

\section{Conclusions}

The chiral limit of the tensor OPE potential has a $1/r^3$ form. In this work, 
I have studied it using techniques developed in atomic physics for solving the
Schr\"odinger equation with inverse-power-law potentials \cite{cava,gao1,gao2}. 
These lead to analytic solutions which are constructed as expansions in Bessel 
functions whose orders satisfy an eigenvalue equation. 
In each channel, there is a critical value of the product of the momentum
and coupling strength above which these eigenvalues become complex. This sets a
limit on the range of energies for which the tensor potential can be treated
perturbatively. These values are the same for both attractive and repulsive
$1/r^3$ potentials.

I have determined the critical dimensionless couplings for the tensor interaction
in low-lying partial waves. In the $^3S_1$--$^3D_1$ and $^3P_0$ channels, the 
corresponding breakdown scales are $\sim m_\pi$ or less. In the other $P$- and
$D$-waves, the scales are of the order of 400~MeV. These results imply that,
for the energies relevant to nuclear physics, OPE should be treated
nonperturbatively in these channels. They explain why Fleming, Mehen and Stewart 
\cite{fms} found that the perturbative KSW treatment breaks down in these cases. 
In contrast, the scales for the higher partial waves all lie well above 1~GeV and 
so perturbation theory should be valid for them.

In the context of the RG, the nonperturbative treatment of OPE can be justified 
if we identify the scale $\lambda_\pi$ controlling its strength as a
low-energy scale. The resulting RG analysis of the $^1S_0$ channel, where only 
the central Yukawa piece contributes, leads to power countings that are similar
to those found for pure short-range interactions. There is a nontrivial fixed point 
which describes systems with strong scattering at low energies. The terms in the 
expansion around this have RG eigenvalues $\rho=-1,0,+1,\cdots$ and so are of order
$Q^d$ where $d=\rho-1=-2,-1,0,\cdots$ \cite{bb1}. This is similar to the power 
counting for the expansion around the nontrivial fixed point for pure short-range 
forces \cite{ksw,bmr}. In both cases the terms in the expansion correspond directly 
to terms in an effective-range expansion.

Here I have used this RG method to study the scaling behaviour in the
spin-triplet channels. This is controlled by the power-law dependence
of the DW's near the origin which, in turn, follows from the singularity
of the long-range potential at the origin. It is the same for attractive and 
repulsive $1/r^3$ potentials. In the $S$-, $P$- and $D$-wave channels, where the 
tensor OPE force needs to be treated nonperturbatively, the expansion around
the trivial fixed point leads to RG eigenvalues $\rho=\frac{1}{2},\frac{3}{2},
\frac{5}{2},\cdots$, corresponding to orders $Q^d$ with $d=-\frac{1}{2},
\frac{1}{2},\frac{3}{2},\cdots$. These are promoted by half an order compared 
with naive dimensional analysis in an $S$-wave and by many more orders in
higher partial waves. This provides a quantitative measure of the effect
observed in Ref.~\cite{ntvk}.

In the $^3S_1$--$^3D_1$ channel we need to expand around the nontrivial fixed point.
The corresponding RG eigenvalues are  $\rho=-\frac{1}{2},\frac{1}{2},\frac{3}{2},
\cdots$ and so there is one relevant perturbation. This is similar to the pure
short-range case, except that the terms in the expansion have been demoted by
half an order.

Finally, in the higher spin-triplet waves, one can treat the tensor OPE potential
perturbatively at low energies. The corresponding scaling behaviour is determined by
the region where the centrifugal barrier dominates, and so the power counting is 
just that given by naive dimensional analysis.

These results show that it is possible to set up a consistent EFT embodying the 
WvK scheme where OPE is treated nonperturbatively in low partial waves. However,
as already remarked, a central element of this is the identification of 
$\lambda_\pi$ as a low-energy scale. Since this scale is built out of quantities
that are treated as high-energy scales in ChPT, this analysis leaves open the 
question of how to make this theory consistent with chiral expansions of other
effective operators, such as those for EM or weak couplings. Further work is needed
to address this.

\section*{Acknowledgments}

I am grateful to the organisers of the ECT* Workshop on ``Nuclear forces and
QCD: never the twain shall meet?" for the opportunity to participate in a very
lively meeting. I thank H. Griesshammer, J. McGovern, A. Nogga,
R. Timmermans and U. van Kolck for helpful discusions, and J. McGovern also for
critical comments on the first draft of this paper.

\end{document}